\documentclass[aps,prd,preprint,superscriptaddress,amsmath,amssymb,showpacs]{revtex4-1}
\usepackage{dcolumn}
\usepackage{graphicx}
\usepackage{float}
\usepackage{physics}
\usepackage[colorlinks=true,allcolors=blue]{hyperref}
\usepackage{mathrsfs}
\usepackage{inputenc}
\usepackage{subfigure}

\newcommand{\be}{\begin{equation}}
\newcommand{\ee}{\end{equation}}
\newcommand{\bea}{\begin{eqnarray}}
\newcommand{\eea}{\end{eqnarray}}

\newcommand{\bln}{\begin{align}}
\newcommand{\eln}{\end{align}}
\newcommand{\bst}{\begin{split}}
\newcommand{\est}{\end{split}}
\newcommand{\bi}{\begin{itemize}}
\newcommand{\ei}{\end{itemize}}
\newcommand{\bn}{\begin{enumerate}}
\newcommand{\en}{\end{enumerate}}

\def\eeq{\end{equation}}

\UseRawInputEncoding
\begin{document}
\title{Complexity Growth in Flavor-Dependent Systems}

\author{Wen-Bin Chang}
\email{changwb@mails.ccnu.edu.cn}
\affiliation{ College of Intelligent Systems Science and Engineering, Hubei Minzu University, Enshi 445000,
People's Republic of China}

\author{Xun Chen}
\email{chenxun@usc.edu.cn}
\affiliation{INFN -- Istituto Nazionale di Fisica Nucleare -- Sezione di Bari, Via Orabona 4, 70125 Bari, Italy}
\affiliation{School of Nuclear Science and Technology, University of South China, Hengyang 421001, People's Republic of China}

\author{Defu Hou}
\email{houdf@mail.ccnu.edu.cn}
\affiliation{Key Laboratory of Quark and Lepton Physics (MOE) and Institute of Particle Physics, Central China Normal University, Wuhan 430079, China}

\date{\today}

\begin{abstract}

In this work, we investigate holographic complexity growth in a flavor-dependent Einstein-Maxwell-Dilaton (EMD) model, where the parameters are determined through machine learning algorithms fitted to lattice QCD equation of state (EoS) and baryon number susceptibility data.
Within the Complexity=Action (CA) conjecture, we introduce a probe string into the bulk geometry and evaluate the time derivative of its Nambu-Goto (NG) action on the Wheeler-DeWitt (WDW) patch as the holographic dual of complexity growth.
Our analysis explores the dependence of complexity growth on string velocity, chemical potential, temperature, and the number of flavors.
Results show maximum complexity growth for stationary strings, decreasing with string velocity. 
At zero chemical potential, complexity growth is largest in the pure gluon system and reduces with the addition of quark flavors.
Increasing temperature and chemical potential consistently enhance complexity growth.
Furthermore, complexity growth exhibits multi-valued behavior in regions corresponding to first-order transitions and single-valued behavior in crossover regimes, indicating that complexity can serve as a probe for phase transitions.

\end{abstract}

\maketitle

\section{Introduction}\label{sec:01_intro}

Motivated by the holographic principle, the connection between quantum information theory and gravitational physics has been explored, particularly through holographic formulations of entanglement entropy and computational complexity \cite{Watrous:2008any,Osborne_2012,PhysRevA.94.040302,Banerjee:2017qti,Takayanagi:2012kg,Hubeny:2007xt,Hashimoto:2018bmb,Susskind:2014rva,Caputa:2017yrh}.
Computational complexity, originally defined in quantum information theory as the minimal number of quantum gates required to prepare a target state from an initial state, is expected to exhibit thermodynamic properties, particularly its analogy with entropy and compliance with the second law of thermodynamics \cite{Bhattacharyya:2018wym,Chapman:2021jbh,Brown:2017jil,Carmi:2017jqz,Fan:2019mbp,Bernamonti:2020bcf}.
Several conjectures have been proposed suggesting that complexity is dual to bulk gravitational observables.
The complexity=volume conjecture suggests that the complexity is proportional to the volume of the Einstein-Rosen bridge \cite{Stanford:2014jda}.
Another proposition is the complexity=action conjecture, which states that complexity is equivalent to the action of a spacetime region known as the WDW patch \cite{Brown:2015bva,Brown:2015lvg}.
Further refinements and generalizations such as the complexity=volume 2.0 and complexity=anything proposals have been introduced, thereby enriching our understanding of complexity \cite{Couch:2016exn,Belin:2021bga,Belin:2022xmt,Jorstad:2023kmq,Myers:2024vve}.
Given the nonlocal nature of holographic complexity, its study should take into account nonlocal operators such as Wilson loop \cite{Moosa:2017yvt,Abad:2017cgl,Fu:2018kcp}.
Nagasaki et al. explored holographic complexity using a nonlocal operator by inserting a fundamental string moving along a great circle in the submanifold, and studied its dependence on string velocity, charge, mass, backreaction, higher-derivative corrections, and Horndeski interactions \cite{Nagasaki:2017kqe,Nagasaki:2018csh,Nagasaki:2019icm,Nagasaki:2022lll,Zhou:2021vsm,Santos:2020xox,Chang:2024muq,Guo:2025bgx,Chang:2025yjw}.

The EMD framework is extensively utilized as a bottom-up holographic approach to quantitatively describe the thermodynamic and transport properties of strongly coupled quark–gluon plasma (QGP) \cite{Gubser:2008ny,DeWolfe:2010he,He:2013qq,Yang:2014bqa,Dudal:2018ztm,Fang:2015ytf,Arefeva:2020vae,Chen:2018vty,Chen:2024ckb,Chen:2024mmd,Zhao:2022uxc,Chen:2020ath,Chen:2019rez,Zhao:2023gur,Hou:2021own,Chen:2025fpd}.
Recent progress involves incorporating machine learning into the EMD framework, which enables the determination of holographic parameters via automatic differentiation and the construction of analytical black hole metrics consistent with the lattice QCD EoS and baryon number susceptibility \cite{Chen:2024ckb,Chen:2024mmd}.
This model incorporates the effects of temperature and chemical potential on QGP and successfully reproduces thermodynamics across different flavor systems, thereby providing a flavor-dependent holographic framework.

Motivated by these developments, this paper investigates the growth of holographic complexity in a flavor-dependent EMD model supported by machine learning.
Within this framework, we introduce a probe string to compute the time derivative of its NG action on the WDW patch, which is interpreted as complexity growth under the CA conjecture.
We systematically analyze how complexity growth depends on string velocity, chemical potential, temperature, and the number of quark flavors in the system.

The paper is organized as follows. 
Sec.~\ref{sec:02} briefly reviews the machine learning supplemented EMD model.
Sec.~\ref{sec:03} details the calculation of the action growth and presents our numerical results. 
Finally, Sec.~\ref{sec:04} provides a summary of our conclusions.

\section{EMD model with machine learning}\label{sec:02}

In this study, the action of 5-dimensional EMD system for describing QCD matter is formulated in the Einstein frame as follows \cite{Yang:2015aia,Dudal:2017max,Chen:2018vty,Chen:2024ckb,Chen:2024mmd}
\begin{equation}
\begin{aligned}
S_{E} & =\frac{1}{16 \pi G_5} \int d^5 x \sqrt{-g}\left[R-\frac{f(\phi)}{4}F^2-\frac{1}{2} \partial_\mu \phi \partial^\mu \phi-V(\phi)\right]. \\
\end{aligned}
\end{equation}
Here, $\phi$ is the dilaton field with potential $V\left(\phi\right)$, $F$ denotes the Maxwell field strength tensor, $f(\phi)$ is the gauge kinetic function, and $G_5$ is the five-dimensional Newton constant.
The explicit forms of $f(\phi)$ and $V\left(\phi\right)$ are derived consistently from the equations of motion.

\begin{figure}
    \centering
    \includegraphics[width=16cm]{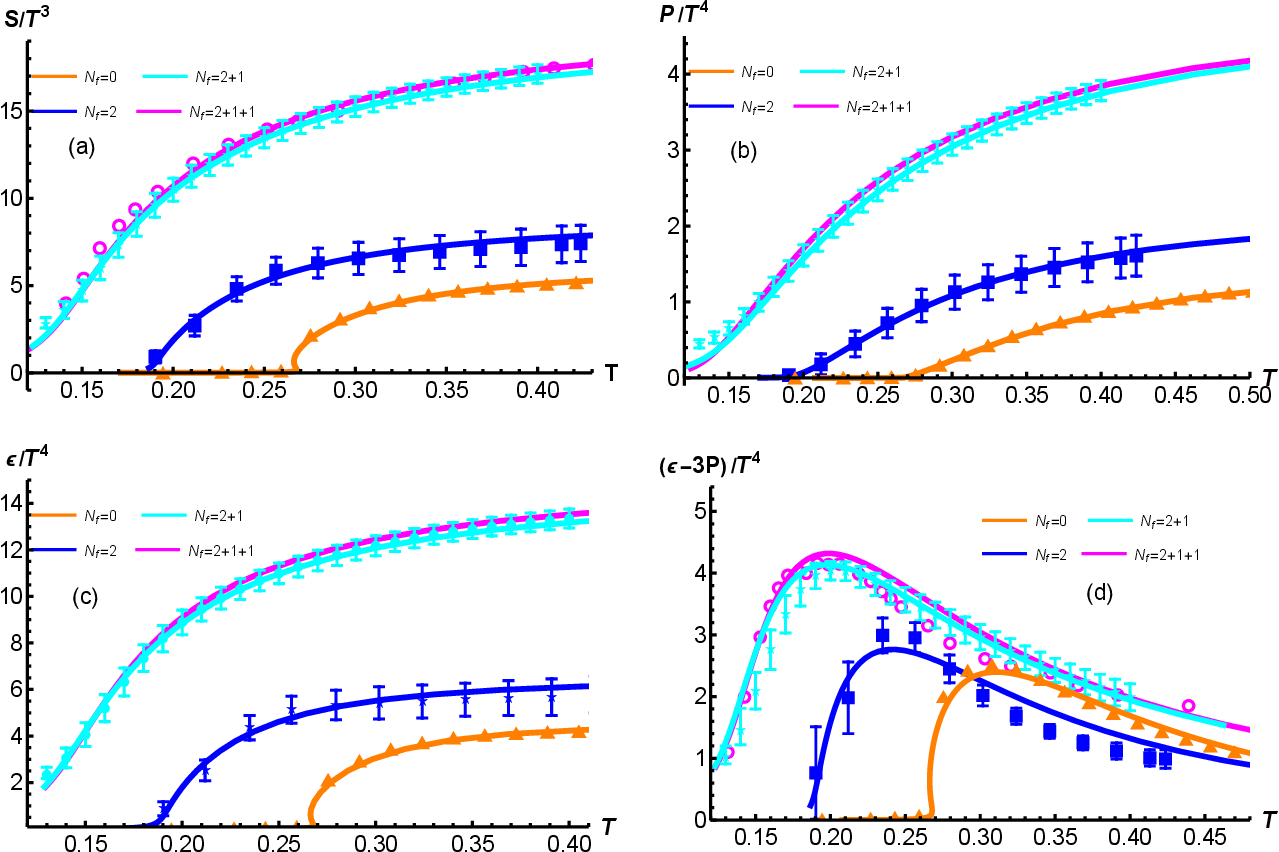}
    \caption{\label{mu0eos}Holographic EoS (lines) compared with lattice QCD results for systems (symbols with error bar) with different flavors \cite{Chen:2024mmd}.}
\end{figure}

We adopt the following metric ansatz
\begin{equation}
d s^2=\frac{L^2 e^{2 A(z)}}{z^2}\left[-g(z) d t^2+\frac{d z^2}{g(z)}+d \vec{x}^2\right],
\end{equation}
where $z$ is the fifth holographic coordinate and the AdS radius $L$ has been set to one.
To obtain analytical solutions for the model, we assume parameterized forms for $f(\phi)$ and $A(z)$
\begin{equation}
A(z)= d \ln(a z^2 + 1) + d \ln(b z^4 + 1), 
\end{equation}
\begin{equation}
f(z)=e^{c z^2-A(z)+k}.
\end{equation}
According to the methodology in \cite{Chen:2024ckb,Chen:2024mmd}, the corresponding solutions are obtained as follows
\begin{equation}
\begin{aligned}
g(z) & =1-\frac{1}{\int_0^{z h} d x x^3 e^{-3 A(x)}}\Big[\int_0^z d x x^3 e^{-3 A(x)}+\frac{2 c \mu^2 e^k}{\left(1-e^{-c z_h^2}\right)^2} \operatorname{det} \mathcal{G}\Big],\\
\phi^{\prime}(z) & =\sqrt{6\left(A^{\prime 2}-A^{\prime \prime}-2 A^{\prime} / z\right)}, \\
A_t(z) & =\mu \frac{e^{-c z^2}-e^{-c z_h^2}}{1-e^{-c z_h^2}}, \\
V(z) & =-3 z^2 g e^{-2 A}\Big[A^{\prime \prime}+A^{\prime}\left(3 A^{\prime}-\frac{6}{z}+\frac{3 g^{\prime}}{2 g}\right)-\frac{1}{z}\left(-\frac{4}{z}+\frac{3 g^{\prime}}{2 g}\right)+\frac{g^{\prime \prime}}{6 g}\Big],
\end{aligned}
\end{equation}
where
\begin{equation}
\operatorname{det} \mathcal{G}=\left|\begin{array}{ll}
\int_0^{z_h} d y y^3 e^{-3 A(y)} & \int_0^{z_h} d y y^3 e^{-3 A(y)-c y^2} \\
\int_{z_h}^z d y y^3 e^{-3 A(y)} & \int_{z_h}^z d y y^3 e^{-3 A(y)-c y^2}
\end{array}\right|.
\end{equation}
The Hawking temperature of the black holes is given by
\begin{equation}
\begin{aligned}
T & =\frac{z_h^3 e^{-3 A\left(z_h\right)}}{4 \pi \int_0^{z_h} d y y^3 e^{-3 A(y)}}\Big[1+\frac{2 c \mu^2 e^k\left(e^{-c z_h^2} \int_0^{z_h} d y y^3 e^{-3 A(y)}-\int_0^{z_h} d y y^3 e^{-3 A(y)} e^{-c y^2}\right)}{(1-e^{-c z_h^2})^2} \Big].
\end{aligned}
\end{equation}
The model contains a six-dimensional parameter space ($a,b,d$ in $A(z)$, $c,k$ in $f(z)$, and $G_5$), whose values for the pure gluon, 2-flavor, 2+1-flavor, and 2+1+1-flavor systems were determined in \cite{Chen:2024mmd} via a machine learning approach trained on lattice QCD data for the EoS and baryon number susceptibility, as shown in Table \ref{table:parameter}.
\begin{table}[htbp]
	\centering
	\begin{tabular}{|c|c|c|c|c|c|c|}
		\hline
		& $a$ & $b$ & $c$ & $d$ & $k$ & $G_5$ \\
		\hline
		$N_f = 0$ & 0 & 0.072 & 0 & -0.584 & 0 & 1.326 \\
        \hline
		$N_f = 2$ & 0.067 & 0.023 & -0.377 & -0.382 & 0 & 0.885 \\
        \hline
        $N_f = 2+1$ & 0.204 & 0.013 & -0.264 & -0.173 & -0.824 & 0.400 \\
        \hline
        $N_f = 2+1+1$ & 0.196 & 0.014 & -0.362 & -0.171 & -0.735 & 0.391 \\
        \hline
	\end{tabular}
\caption{Parameters obtained by machine learning for the pure gluon, 2-flavor, 2+1-flavor, and 2+1+1-flavor systems, respectively.}
\label{table:parameter}
\end{table}
With the machine learning determined parameter sets, our holographic model closely fits the EoS for systems with different flavors when compared with lattice QCD data, as illustrated in Fig.~\ref{mu0eos} \cite{Chen:2024mmd}.

\section{EVALUATION OF THE COMPLEXITY GROWTH}\label{sec:03}
Based on the methodology of \cite{Nagasaki:2017kqe}, we investigate the growth of the NG action in this machine learning supported EMD system with a probe string, which provides a dual representation of complexity growth in accordance with the CA conjecture.
In the string frame, where $A_{s}(z) = A(z) + \sqrt{1/6}\phi(z)$ and $b(z) = e^{2As}$, the NG action is computed by parameterizing the string worldsheet with the coordinates
\begin{equation}
\label{eqs}
\ t=\tau,\  \ r=\sigma, \ \phi=v\tau+\xi(\sigma).
\end{equation}
In this setting, the function $\xi(\sigma)$ determines the shape of the string, and $\upsilon$ denotes its velocity with respect to the black hole.
To compute the time derivative of the NG action, we integrate the square root of the determinant of the induced metric over the WDW patch
\begin{equation}
\label{eqt}
\frac{d S_{N G}}{d t}=T_{s} \int_{z_{h}}^{\infty} d \sigma \sqrt{-g_{\text {ind }}(\sigma)}=T_{s} \int_{z_{h}}^{\infty} d \sigma \frac{b(z)\sqrt{\xi^{\prime 2} g(z) - \frac{v^2}{g(z)} + 1}}{z^2}.
\end{equation}
Here, $T_{s}$ is the fundamental string tension.
The equation of motion for $\xi$ is derived by varying the above action
\begin{equation}
\label{equ}
\ \Pi_{\xi}=\frac{b(z) \xi' g(z)}{z^2 \sqrt{\xi^{\prime 2} g(z) - \frac{v^2}{g(z)} + 1}},
\end{equation}
with $\Pi_{\xi}$ representing the integration constant.
We can therefore resolve $\xi'^{2}$ in the form of
\begin{equation}
\label{eqv}
\xi'^{2}=\frac{\Pi^2_{\xi} z^4 \left(g(z) - v^2\right)}{g(z)^2 \left(b(z)^2 g(z) - \Pi^2_{\xi} z^4\right)}.
\end{equation}
To ensure the above expression yields real values, the numerator and the denominator must change sign simultaneously, which determines the critical point as
\begin{equation}
\label{eqw}
g(z_c)=v^2,
\end{equation}
and
\begin{equation}
\label{eqx}
b(z_c)^2 g(z_c)= \Pi^2_{\xi} z_c^4.
\end{equation}
Finally, the growth of the NG action is obtained by substituting (\ref{eqv}), (\ref{eqw}), and (\ref{eqx}) into (\ref{eqt}), which yields
\begin{equation}
\label{eqfn}
\frac{d S_{N G}}{d t}=T_s \int_{z_h}^{\infty} d \sigma\sqrt{\frac{b(z)^4 z_c^4 \left(g(z) - v^2\right)}{z^4 \left(b(z)^2 g(z) z_c^4 - b(z_c)^2 v^2 z^4\right)}}.
\end{equation}

In the following, the action growth for different flavors characterized by the parameter sets in Table \ref{table:parameter} is investigated by numerically integrating (\ref{eqfn}) and the results are presented in the subsequent figures.
    
\begin{figure}[H]
    \centering
      \setlength{\abovecaptionskip}{-0.1cm}
    \includegraphics[width=7cm]{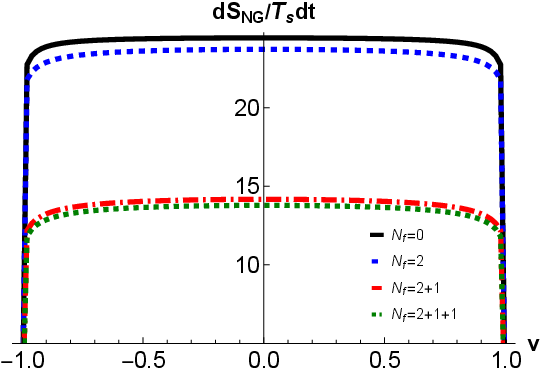}
    \caption{\label{fig1} Action growth versus string velocity for different flavors at $T= 0.6GeV$.}
\end{figure}

Fig.~\ref{fig1} illustrates the dependence of action growth on string velocity at zero chemical potential for different flavors.
For all flavors, Fig.~\ref{fig1} shows that the action growth is maximized when the string is stationary $(\upsilon=0)$.
The number of flavors in the system has a significant impact on the action growth, which is maximized in the pure gluon system and subsequently decreases as the number of flavors increases.
Furthermore, the difference between the 2+1-flavor and 2+1+1-flavor systems is relatively small, whereas they both differ significantly from the 2-flavor system.

\begin{figure}[H]
    \centering
    \setlength{\abovecaptionskip}{-0.1cm}
    \subfigure{
        \includegraphics[width=5cm]{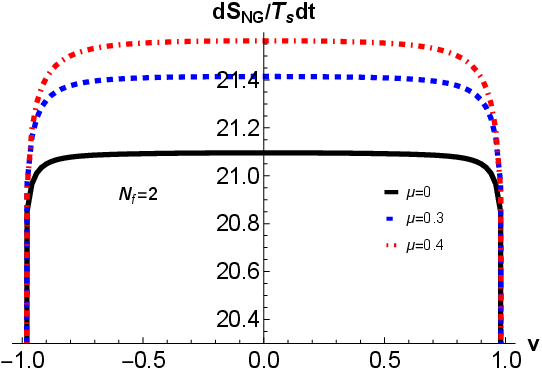}
    }
    \subfigure{
        \includegraphics[width=5cm]{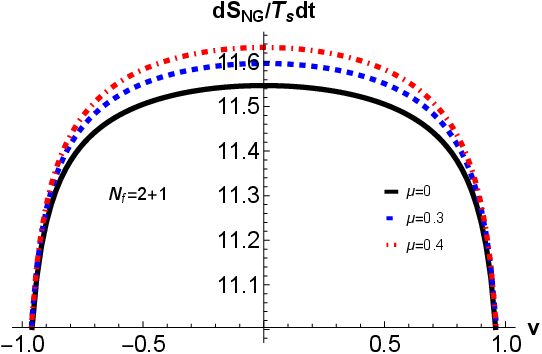}
    }
    \subfigure{
        \includegraphics[width=5cm]{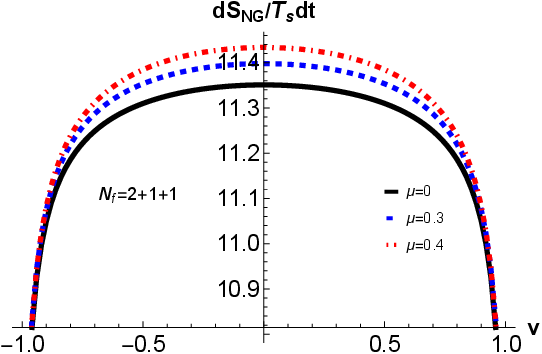}
    }
    \caption{\label{fig2} Action growth versus string velocity for different chemical potentials at $T= 0.2GeV$ for the $N_{f} = 2$, $N_{f} = 2+1$, and $N_{f} = 2+1+1$ flavor systems.}
\end{figure}

Fig.~\ref{fig2} illustrates the influence of the chemical potential on the action growth for various flavor systems.
In both cases, increasing the string velocity reduces the action growth. 
This result is consistent with the findings in Fig.~\ref{fig1} and implies that string motion tends to suppress action growth.
For $N_{f} = 2$, $N_{f} = 2+1$, and $N_{f} = 2+1+1$ systems, increasing the chemical potential $\mu$ consistently enhances the action growth, thereby accelerating the growth of complexity.
This observation is consistent with the findings previously reported in \cite{Chang:2024muq}.

\begin{figure}[H]
    \centering
      \setlength{\abovecaptionskip}{-0.1cm}
    \includegraphics[width=7cm]{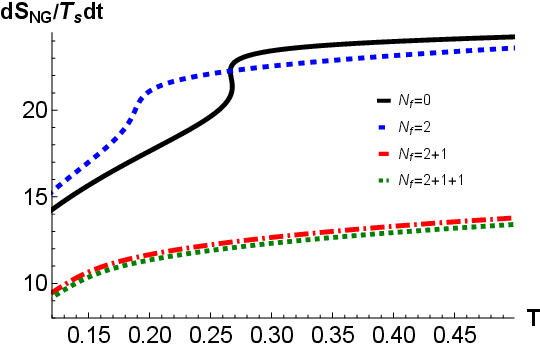}
    \caption{\label{fig3} Action growth versus temperature at vanishing chemical potential for different flavors with $\upsilon=0$.}
\end{figure}

Fig.~\ref{fig3} depicts how action growth changes as a function of temperature at vanishing chemical potential for different flavors.
In Fig.~\ref{fig3}, the pure glue system exhibits a multi-valued behavior around the critical temperature $T_c \approx 0.265 \, \text{GeV}$, indicating a first-order phase transition in the system.
For the systems with flavors $N_{f} = 2$, $N_{f} = 2+1$, and $N_{f} = 2+1+1$, the action growth is a single-valued function of temperature, indicating that the transition in these cases is a crossover.

\begin{figure}[H]
    \centering
    \setlength{\abovecaptionskip}{-0.1cm}
    \subfigure{
        \includegraphics[width=5cm]{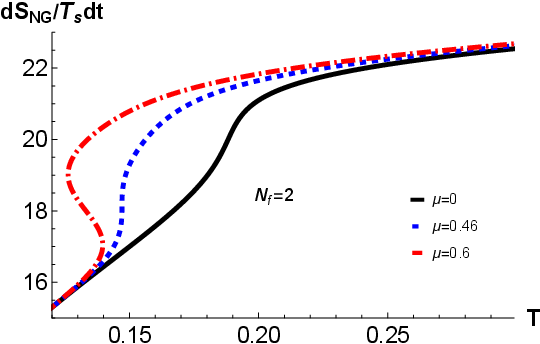}
    }
    \subfigure{
        \includegraphics[width=5cm]{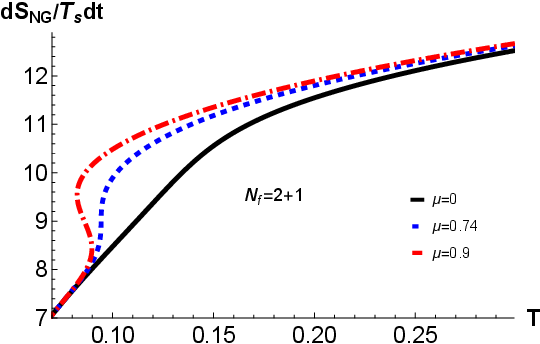}
    }
    \subfigure{
        \includegraphics[width=5cm]{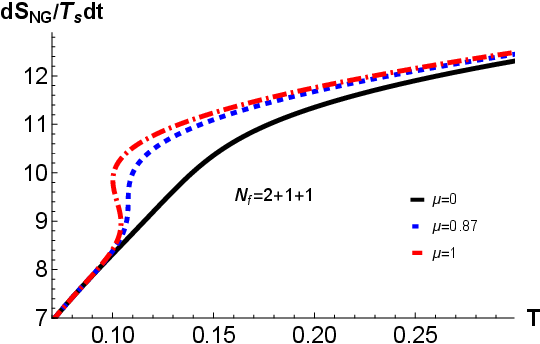}
    }
    \caption{\label{fig4} Action growth versus temperature at finite chemical potential for the $N_{f} = 2$, $N_{f} = 2+1$, and $N_{f} = 2+1+1$ flavor systems with $\upsilon=0$.}
\end{figure}

Fig.~\ref{fig4} illustrates the temperature dependence of action growth at finite chemical potential.
As shown in Fig.~\ref{fig4}, the action growth is a single-valued function of temperature at low chemical potential in 2-flavor, 2+1-flavor, and 2+1+1-flavor systems.
At high chemical potential, the action growth exhibits a multi-valued behavior, which means a first-order phase transition is present.
Furthermore, excluding the critical region, the action growth shows a consistent increase with temperature.

\section{CONCLUSION}\label{sec:04}
In this work, we have investigated the holographic complexity growth in a flavor-dependent EMD model.
The model parameters for pure gluon, 2-flavor, 2+1-flavor, and 2+1+1-flavor systems were determined via machine learning algorithms trained on lattice QCD data for the EoS and baryon number susceptibility.
We evaluated the time derivative of the NG action for a probe string in the bulk, which is interpreted as the holographic dual of complexity growth within the framework of the CA conjecture.
As a result, our study systematically investigates the dependence of complexity growth on string velocity, chemical potential, temperature, and the number of flavors.
Our main findings can be summarized as follows.

For all flavor systems, the complexity growth, represented by the NG action growth, reaches its maximum when the probe string is at rest and decreases monotonically with increasing string velocity.
This observed feature accords with findings from previous studies and may indicate a universal property of holographic complexity \cite{Nagasaki:2018csh}.
At a fixed temperature and vanishing chemical potential, the complexity growth shows a significant dependence on the number of flavors, being largest in the pure gluon system and decreasing as the number of flavors increases.
This indicates that the inclusion of $u$ and $d$ quarks decreases the complexity growth, which is further reduced by the addition of $s$ and $c$ quarks.
For the $N_{f} = 2$, $N_{f} = 2+1$, and $N_{f} = 2+1+1$ flavor systems, increasing the chemical potential consistently leads to an enhancement of complexity growth.
Since higher chemical potential corresponds to larger quark number density, our findings suggest that denser environments promote the complexity growth.
Notably, we observed that at vanishing chemical potential, the complexity growth is multi-valued for the pure gluon system, indicating a first-order phase transition, but remains single-valued in the $N_{f} = 2$, $N_{f} = 2+1$, and $N_{f} = 2+1+1$ flavor systems, consistent with crossover transitions. At finite chemical potential, it is single-valued at low $\mu$ but becomes multi-valued at high $\mu$, signaling the emergence of a first-order phase transition.
These results suggest complexity growth may serve as an effective tool for characterizing phase transitions.

In future work, it will be valuable to extend this investigation to systems with finite magnetic fields or rotation to construct more realistic holographic models.

\section*{Acknowledgments}

Defu Hou's research is supported in part by the National Key Research and Development Program of China under Contract No. 2022YFA1604900. Additionally, Defu Hou receives partial support from the National Natural Science Foundation of China (NSFC) under Grant No.12435009, and No. 12275104. 
Xun Chen the National Natural Science Foundation of China (NSFC) under Grant No. 12405154, and the European Union -- Next Generation EU through the research grant number P2022Z4P4B ``SOPHYA - Sustainable Optimised PHYsics Algorithms: fundamental physics to build an advanced society'' under the program PRIN 2022 PNRR of the Italian Ministero dell'Universit\`a e Ricerca (MUR).
Wen-Bin Chang is supported by the Ph.D. Research Startup Project at Hubei Minzu University (Project No. RZ2500000857). 

\bibliography{ref}
\end{document}